\documentclass[12pt]{article} 
\usepackage{graphicx}
\usepackage{epsfig}
\usepackage{psfig}
\usepackage{nature}
\def\gtap{\mathrel{ \rlap{\raise 0.511ex \hbox{$>$}}{\lower 0.511ex
   \hbox{$\sim$}}}} 
\def\ltap{\mathrel{ \rlap{\raise 0.511ex
   \hbox{$<$}}{\lower 0.511ex \hbox{$\sim$}}}} 
\newcommand{\beq}{\begin{equation}}

\newcommand{\eeq}{\end{equation}}
\newcommand{\bea}{\begin{eqnarray}}
\newcommand{\eea}{\end{eqnarray}}

\begin{document}

\title{\bf Long and short gamma-ray bursts, and the pulsar kicks}

\author{ Alexander Kusenko$^{*\dag}$ and Dmitry V. Semikoz$^{*\ddag}$ \\ 
\small $^*$ Department of Physics and
  Astronomy, UCLA, Los Angeles, CA 90095-1547\\
\small $^\dag$ RIKEN BNL Research Center, Brookhaven National
Laboratory, Upton, NY 11973 \\
\small $^\ddag$ INR RAS,
60th October Anniversary prospect 7a, 117312, Moscow, Russia}

\date{} 
\maketitle

\begin{abstract}

One of the mysteries that surround gamma-ray bursts (GRB) is the origin of
two classes of events: long and short GRB\cite{short-long}.  The short GRB
are similar to the first second of a long GRB\cite{griz03a}.  We
suggest that the short bursts are interrupted long bursts, we point out
a plausible mechanism for the interruption, and we explain the observed
time scales. 
The supernova-like central engine may contain a
neutron star or a black hole surrounded by an accretion disk and jets.  In
the case of a neutron star, the same mechanism that is responsible for the
pulsar kicks can disrupt the central engine, thus producing an interrupted,
short GRB.  If a black hole is produced, the absence of the kick ensures
the long duration of the GRB.  The time delay of the kick and the
absence of the kick for a black hole are natural consequences of a model
based on neutrino oscillations\cite{fkmp}.


\end{abstract}

Observational evidence suggests that gamma-ray bursts are supernova-like
phenomena\cite{GRB-SN}.  This evidence is stronger for the long GRB, which
last longer than a few seconds and for which the optical afterglows have
been detected.  It is now established that the long GRB occur in star
forming regions.  The central engine is probably a sort of a supernova
explosion, which is different from ordinary supernovae in that a large
fraction of energy is concentrated in highly relativistic jets.  On the
other hand, for short GRB, the lack of optical afterglows leaves open
several other possibilities.  For example, neutron star and black hole
mergers are probably not ruled out\cite{ns-ns}.

In this {\em Letter} we will argue that both  long and short GRB can
originate from the same supernova type central engine that produces either
a black hole or a neutron star, respectively.  The main reason why the short
bursts are shorter is that the jet, pointing originally along the line of
sight, is disrupted by the same dynamical mechanism that is responsible for
the ``pulsar kicks''.  A model for such a mechanism\cite{fkmp} based on
neutrino oscillations predicts a delay on the time scale of order of a few
seconds, which can naturally explain the durations of the interrupted
bursts.  The same kick mechanism does not act on a black hole\cite{fkmp}.
Thus, if the central engine is a black hole, the jet exists and maintains its
direction along the line of sight for (much) longer than a second, which
results in a long GRB.  We suggest, therefore, that the long GRB come from
central engines containing black holes, while the short ones originate from
those with neutron stars.

The pulsar kick may also be the origin of unusually high rotational 
velocities of pulsars at birth\cite{sp}.  The kick is not expected to be a
central force.  Hence, it simultaneously kicks the pulsar and spins it up.
At the time of the kick both the position and the angular momentum of the
neutron star change dramatically.  

Since the accretion disk and the jet are gravitationally bound to the
neutron star, the pulsar kick can disrupt the jet entirely.  An average
pulsar velocity is 200-500 km/s, although about 15\% of pulsars have
velocities in excess of 1000 km/s\cite{velocities}.  The neutron star is
strongly gravitationally bound to a part of the accretion disk of the size,
roughly, 10 neutron star radii, $D \sim 10 R_{_{\rm NS}} \approx
150$~km. Therefore, it takes from 0.1 to 1 second after the kick for the
neutron star to leave the central region of the accretion disk.  This can
either terminate the jet entirely, or can change its direction.  The kick
changes the angular momentum of neutron star, which affects the total
angular momentum of the system and, thus, the direction of the jet.  In
either case, whether the jet is disrupted entirely or it is kicked off the
line of sight, the kick terminates the GRB.

The two essential elements of this picture, illustrated in Fig.1, are the
presence of a compact star in the central engine and the existence of a
pulsar kick mechanism which acts on a neutron star within one second of
the onset, and which does not act on a black hole. 
We will first discuss the phenomenological consequences of such a
mechanism, and then we will argue that a microscopic model based on
neutrino oscillations\cite{fkmp} is suitable.

This simple explanation can account for the following observational features of
GRB:  

\begin{itemize}
\item short-time variability on the millisecond time scale for all bursts 
\item temporal and spectral similarities between the short bursts and the
  first seconds of the long bursts  
\item lack of the optical afterglows following the short GRB
\item approximate fluence-duration proportionality 
\item characteristic time scale of about $1\,$s separating the two classes
of bursts
\item supershort bursts
\end{itemize}
We will discuss each of these items below.

{\em Short-time variability} is a natural consequence of a small size of
the compact star at the center of the engine.  The observed gamma rays are
generated in the jet by some dynamics that may involve shocks and flow
irregularities.  However, the millisecond time scale in the temporal
profile of bursts is suggestive of a small size dynamics that modulates the
jet at its origin.  Causality forces one to associate the millisecond time
intervals with length scales of the order of a kilometer.  Hence, a compact
central object is called for.  In addition, the smaller the central engine,
the stronger is the field of gravity in the region where the jet
originates.  This can help accelerate the accreting material to higher
speeds before it is ejected out into the jet. Given the tight energetic
constraints on the central engine, it is desirable to extract as much
energy as possible in the form of a narrow jet.  Therefore, the presence of
a compact star in the central engine is definitely a desirable feature of
the model.  An improvement in temporal sensitivity of observations in the
millisecond range, in combination with improved theoretical modeling, may
help distinguish black holes from neutron stars by measuring short-time
variability, which may be different for those two cases.
  
\begin{figure}
{\epsfysize=4cm 
\epsfbox{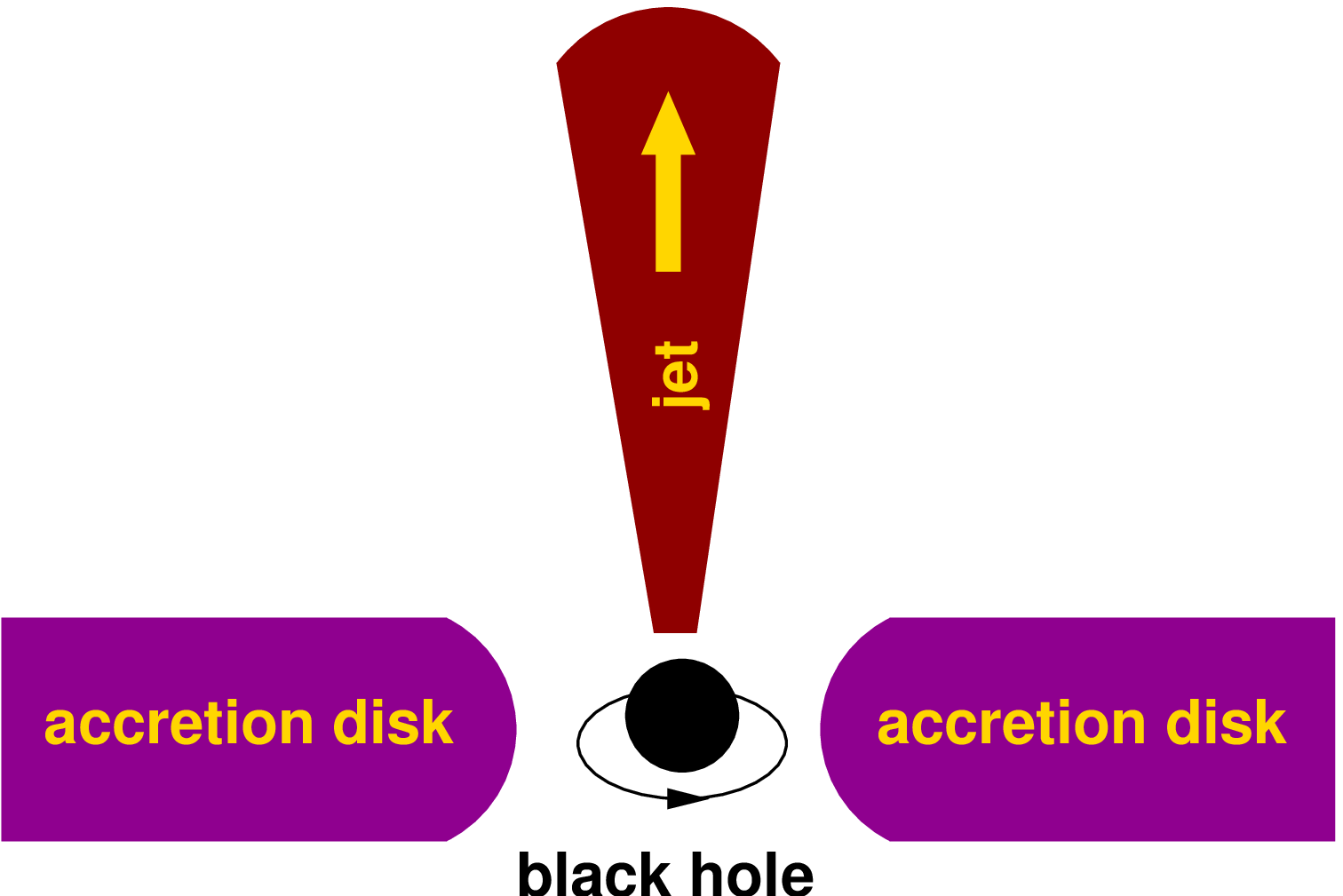}} \hfill
{\epsfysize=4cm 
\epsfbox{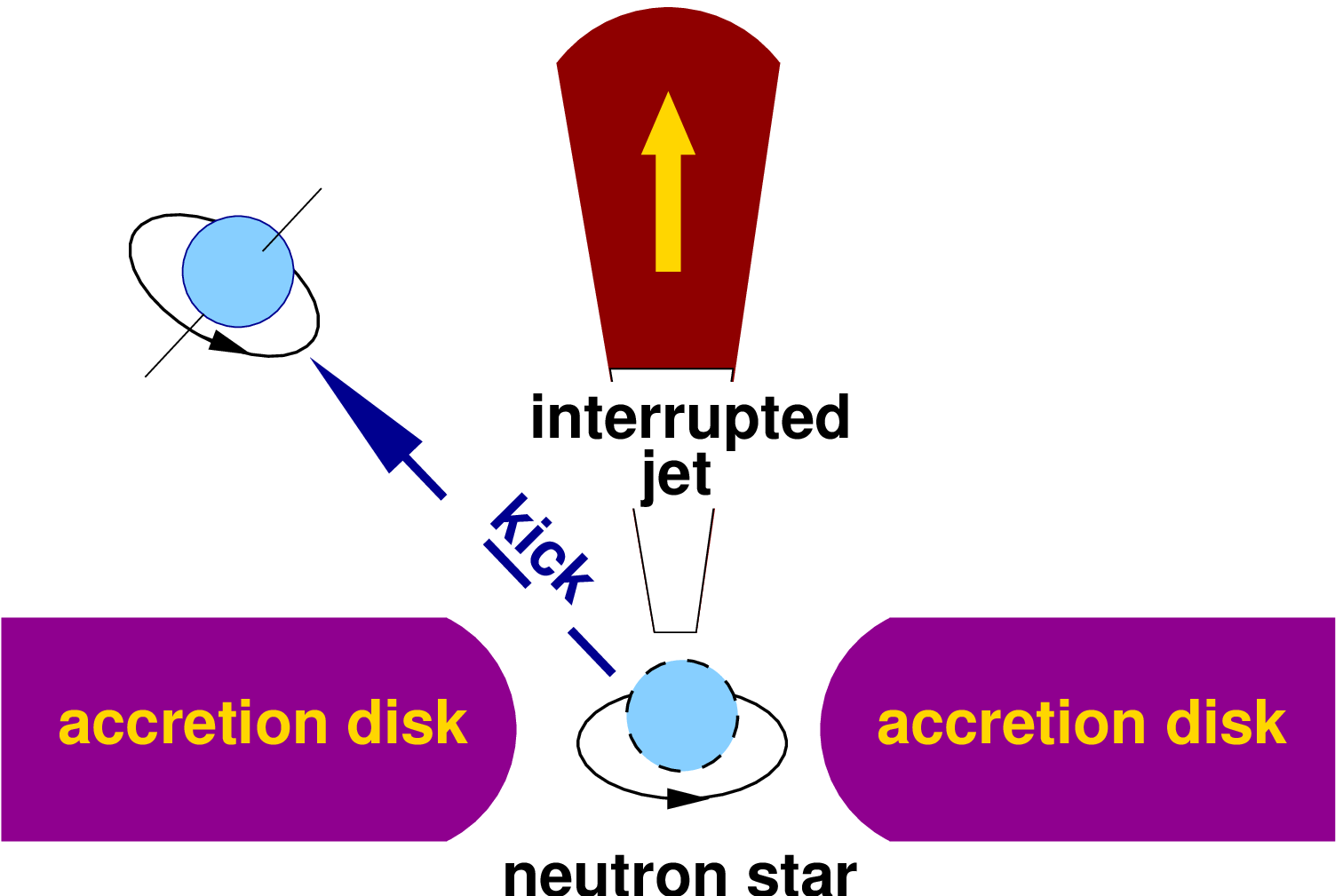}
}
\caption{ \small 
The central engine may contain a black hole or a neutron star.
In the case of a black hole, the jet continues uninterrupted for tens of
seconds. This is long GRB.  In the case of a neutron star, about a second
after the onset of the GRB, the pulsar kick (due to neutrinos\cite{fkmp},
or some other mechanism) displaces the neutron star and changes its angular
momentum, disrupting the central engine.  This produces a short, interrupted
gamma-ray burst with similar characteristics to the first few seconds of a
long burst\cite{griz03a}. }
\label{figure}
\end{figure}

{\em Temporal similarities between the short and long bursts} are to be
expected if the short bursts are merely the interrupted long bursts.  This
is strongly supported by recent observations\cite{griz03a}, which
demonstrate that the first seconds of the long bursts are almost identical
to the short bursts.  Improved future analyses may reveal a softening of the
short GRB toward the end, when the central engine is disrupted.

{\em The lack of optical afterglows for the short bursts} may be the
consequence of a disruption of the central engine and the jet by the kick
force. The isotropic emission of light in a supernova is insufficient to
produce an observable afterglow at large red shift.  Thus, after the jet is
disrupted, one does not expect an afterglow of observable luminosity.  In
the black hole case, the jet remains stable for the entire duration of the
(long) GRB, and the same jet can be detected at a later time as an
afterglow. However, an early optical afterglow for a short GRB is also
possible if it results from an interaction of the 
interrupted jet with ambient matter.
 
{\em The approximate fluence-duration
proportionality}\cite{fluence-duration} is also explained in our model
because the short bursts are supposed to be the interrupted long ones.  In
the absence of the interruption by the kick force, the short burst would
have continued and would have produced the total luminosity typical of a
long burst.  The disruption of the jet ends the emission of energy in the
direction of the observer.

The fluence -- duration power law is softer for the short GRB then it is
for the long ones\cite{fluence-duration}.  This observation fits in our
model because the black holes can make more powerful engines than the
neutron stars.  This is because the black holes can have a larger mass and
a smaller size simultaneously.

{\em The characteristic time scale of order 1~s} arises naturally in a
model that explains the pulsar kicks by neutrino oscillations\cite{fkmp}.
There are actually two viable mechanisms, based on the
resonant\cite{ks97} and non-resonant\cite{fkmp} oscillations,
respectively.  The unknown neutrino parameters determine which of the two
(mutually exclusive) mechanisms is a work.  In the case of resonant
oscillations, the force is applied from the onset of the neutron star
cooling, with no delay.  However, for different neutrino parameters, the
off-resonant mechanism\cite{fkmp} is operative, and the kick starts only
after some time $\tau \sim 1\,s$, which is a natural time scale to explain
the difference between the short and the long GRB.  This pulsar kick
mechanism requires that a singlet fermion, having a keV mass and mixed with
the electron neutrino, be present in the neutrino spectrum.
Serendipitously, the same singlet neutrino can be the dark matter in the
universe\cite{fkmp,dm}.  The kick begins after the effective matter
potential for neutrino oscillations is reduced by neutrino conversions.
The time it takes\cite{fkmp} before the start of the kick is estimated to be 
\begin{eqnarray}
\label{timeeqoffres}
 \tau  & \simeq &
\frac{4 \sqrt{2} \pi^2 m_n}{G_{\!\!_F}^3 \rho}
\frac{ (V_m^{(0)})^3}{(\Delta m^2)^2 \sin^2 2 \theta } \frac{1}{\mu^3}
\\
& \sim & 1 {\rm s} \, 
\left ( \frac{10^{-8}}{\sin^2 2 \theta} \right ) 
\left (\frac{50 \mathrm{MeV}}{\mu} \right )^3 
\left ( \frac{ 
10 \, \mathrm{keV}^2
}{\Delta m^2
} \right )^2, 
\nonumber   
\label{timeeqoffresnumerical}
\end{eqnarray}
where $\rho$ and $\mu$ are the density and the electron chemical potential
inside the neutron star, $\Delta m^2$ is the difference in neutrino masses
squared, $\theta$ is the singlet-active neutrino mixing angle, and $
V_m^{(0)}\sim 0.1 \, {\rm eV}$ is the initial value of the matter potential.
The mixing $\sin^2 2\theta \sim 10^{-11}-10^{-7}$ for a $1-10\,$keV singlet
neutrino is consistent with dark matter\cite{dm}.  Clearly, the one-second
time scale is a natural prediction of this mechanism.

Even if there is no delay, and the pulsar kick occurs at the time of the
formation of the jet, the observed pulsar velocities of the order of a few
hundred km/s can explain the the $\sim1\,$s time scale.  Indeed, it may be
necessary to remove the pulsar at least a distance $D\sim 10 R_{_{NS}} \sim
150\,$km away from its original position to disrupt the jet.  Thus, the jet
may remain intact for the time $\sim D/v$, where $v$ is the pulsar
velocity.  The range of time delays that arise this way is determined by
the distribution of pulsar velocities\cite{velocities}. Since most pulsars
move faster than 100 km/s, we do not expect a burst that lasts much longer
than a second as long as the engine is powered by a neutron star.

{\em Supershort bursts}. Gamma-ray bursts with a very short duration
$\tau<0.1$ second are sometimes considered as a separate class of
GRB\cite{cline}.  It is possible that such short bursts appear because a
jet, which was not pointing at Earth initially, changes its direction and
quickly passes through the line of sight when the neutron star is kicked.
Alternatively, the supershort bursts may correspond to the neutron stars
with the highest velocities, which move out of the central region in much
less than a second. 

Our model can, in principle, relate the numbers of the long and
the short gamma-ray bursts to the birth rates of black holes and neutron
stars, respectively.  Simple estimates of the integrated galactic birth
rates for compact stars\cite{st} suggest that neutron stars are born at 
two or three times the rate of black holes.  However, neutron
stars and black holes are born from progenitors of different masses.  This
may affect the angular size of the jet, the brightness, and, ultimately, the
detectability of the GRB.  Therefore, we cannot relate the numbers of the
short and long GRB directly to the birth rates of compact objects. 

Our model agrees with non-observation of {\em late} optical afterglows for 
short GRB. If such afterglows will be observed in the future,
interpretation of short GRB as interrupted jets will become less convincing.

To summarize, we have proposed that the long GRB originate from
supernova-like explosions that result in a formation of a black hole, while
the short GRB originate in the same way, except a neutron star forms
instead of a black hole.  The neutron stars undergo the same dynamical
impacts that cause the pulsar kicks.  As a result, the momentum and the
angular momentum of the neutron star and of the gravitationally bound
accretion disk change dramatically.  This impact either disrupts the jet
altogether or changes its direction away from the line of sight.  Central
engines that contain neutron stars can, therefore, produce short GRB.  At
the same time, a central engine with a black hole, which is not a subject
to the kick force (at least as long as the pulsar kick is due to neutrino
oscillations), would produce a long GRB.  At least one plausible model for
the pulsar kick mechanism\cite{fkmp} predicts that the kick force would be
delayed by a period of time of order one second.  However, regardless of
the microscopic mechanism, since most pulsars have velocities in excess of
100 km/s\cite{velocities}, the pulsar kick should remove the neutron star
from the central region within about a second, which 
should destabilize the jet.  Our model can be generalized to include any
kick mechanism that acts on a neutron star but does not act on a black
hole.  This can be the explanation for the time separation between the two
populations of bursts.

{\bf Acknowledgments} This work was supported in part by the
U.S. Department of Energy and NASA grants. 

{\bf Competing interests statement} The authors declare that they have no
competing financial interests.  

{\bf Correspondence} to D.S. (email: semikoz@physics.ucla.edu).  

\end{document}